\documentstyle[pre,aps,epsf,epsfig]{revtex} 
\begin{document}
\draft
\twocolumn[\hsize\textwidth\columnwidth\hsize\csname@twocolumnfalse\endcsname
%
\title{A Spring-block Model for Barkhausen Noise }
%
\author{K. Kov\'acs$^1$, Y. Brechet$^2$ and Z. N\'eda$^{1,3}$}
\address{ $^1$Babe\c{s}-Bolyai University, Dept. of Physics, 
          str. Kog\u{a}lniceanu 1, RO-400084 Cluj-Napoca, Romania \\
          $^2$ Institut National Polytecnique de Grenoble, ENSEEG-LTPCM, 38402 Saint Martin d'Heres, France \\
          $^3$ Centro de Fisica do Porto, 4099-002 Porto, Portugal\\
          E-mail: zneda@phys.ubbcluj.ro}

\maketitle

\begin{abstract}

A simple mechanical spring-block model is introduced for studying
magnetization phenomena and in particularly the Barkhausen noise. The model
captures and reproduces the accepted microscopic picture
of domain wall movement and pinning. Computer simulations suggest that this model
is able to reproduce the main characteristics of hysteresis loops and Barkhausen jumps.
In the thermodynamic limit the statistics of the obtained Barkhausen jumps
follows several scaling laws, in qualitative agreement with the 
experimental results. The simplicity of the model and the invoked
mechanical analogies makes it attractive for computer simulations and pedagogical purposes.

\end{abstract}

\vspace{2pc}
]

\vspace{1cm}

\section{Introduction}

Barkhausen noise (BN) belongs to the family of the so-called crackling noises \cite{set}.  It
appears as a series of discrete and abrupt jumps in the magnetization when a
ferromagnetic sample is placed under varying external magnetic field. It is believed
that the BN is a consequence of the fast movement of domain walls between pinning centers,
which are either defects or impurities in the ferromagnetic sample. The present paper
intends to introduce a simple and successful mechanical model for describing this classic
magnetization phenomenon. The simplicity of the model and the invoked mechanical analogies
make this model attractive for computer simulation studies and pedagogical purposes.

The Barkhausen phenomenon is interesting from several points of view. From a practical side, by
measuring the BN there is a possibility for non-destructive and non-invasive material
testing and control. On the other hand, from a pure conceptual viewpoint, by studying
the BN one might reach a better understanding of the complex dynamics of domain walls
during magnetization processes.

Since its discovery (1917) BN has been intensively studied. Numerous measurements were
done to clarify the statistical properties of the BN \cite{spas,plew,obrien}.  Regarding
the nature of the Barkhausen noise (white noise, $1/f$ noise, or $1/{f}^{2}$ noise) even the experimental
results are incoherent with each other. The most extensive measurements and data analysis
were performed by Spasojevic et al. \cite{spas} with a commercial VITROVAC 6025-X metal glass
(quasi 2D) sample. After performing the statistical and numerical calculations, they have
found: (i) power-law type distribution for signal duration with scaling exponent $-2.22$,
(ii) power-law behavior for signal area with scaling exponent $-1.77$, and (iii) power-law
type power spectrum with scaling exponent $-1.6$ to $-1.7$. From here they concluded that BN is
not pure $1/f$, nor $1/{f}^{2}$ (Brownian) type noise, but something between these two. Plewka,
et al. \cite{plew} performed measurements (and calculations) with a similar experimental
setup on an amorphous ribbon in an open magnetic circuit. They obtained instead a value
around $-0.9$ for the scaling exponent of the power spectrum. From this result they concluded
that BN is typically $1/f$ noise. O'Brien and Weissman \cite{obrien} performed measurements with a
SQUID magnetometer on an amorphous iron-based metallic alloy (2605TCA) and they suggested
that BN is much closer to a white noise than a $1/f$ noise and differs sharply from most
typical $1/f$ noises.

BN received a special attention in the context of self-organized criticality.
Self -organized criticality (SOC) is a term used for a class of complex phenomena where
non-equilibrium broadband noise in driven systems reflects a type of self-organization,
producing states with power-law correlations closely analogous to critical phenomena \cite{btw}.
Some of the ingredients of SOC were known to be potentially relevant to BN. In some cases,
magnetization changes have been directly observed to occur via avalanche process in the
domain topology \cite{bab}. These avalanches exhibit some scaling effects, at least over a
narrow range of parameters, and their behavior has been described by a SOC model \cite{bakfl,cote}.
There are however other approaches that put under doubt the relevance of the SOC concept
to BN. O'Brien and Weissman \cite{obrien} for example argued that the presumed $1/f$ nature of BN
and the observed power-law distributions are not necessarily evidences of SOC, but rather
the consequences of the scaling properties of quenched disorder in the material.

Many conceptually different models were elaborated to explain BN and its scaling properties.
Without the intention of making a complete review, here we will mention only a few
selected theoretical approaches. Alessandro et al. \cite{abbm} proposed a single degree of
freedom model (ABBM model) that considers the motion of a single domain wall in a
spatially rough coercive field created by the defects. They concluded that a mean-field
approximation is adequate, and found power-law behavior for the Barkhausen pulse size
distribution. Another model \cite{durin} which is strongly related to the previous one considers
the motion of a single flexible domain wall in an uncorrelated disordered medium. This
approach leads to a power-law distribution of the avalanche sizes (exponent $-1.5$) and
durations (exponent $-2$), and yields an exponent $-2$ for the scaling of the power spectrum.
Perkovic, Dahmen and Sethna \cite{perk} described the BN in terms of avalanches near
a critical point. They used the zero-temperature random field Ising model (RFIM), in which the
effect of the pinning centers was taken into account as a normally distributed local
random field. This model was able to account for the power-laws characteristic for the
distribution of avalanche sizes, signal area and signal duration. Another theoretical
attempt by O. Narayan \cite{nara} considers a multiple degrees of freedom model, studying the
relaxational dynamics of a single domain wall in a two-dimensional Ising system. This
model yields a  power-law with critical exponent $-1.5$ for the power spectrum. The model
predicts that in other dimensions different critical exponents are expected, for example
in one-dimension the critical exponent for the power spectrum should be zero. This result
is in contradiction with the prediction of the mean-field approximation for the single
degree of freedom model \cite{abbm}, which predicts the value $-2$ for the exponent,
independently of the dimensionality of the model.

Despite of the numerous complex models and conceptually different approaches, presently
none of them is able to face realistically all the experimentally observed features of
the Barkhausen phenomenon. The spring-block model which will be introduced in this paper offers
an elegant alternative. As it will be shown in the following, this model has the potential
to explain within a relatively simple manner most of the experimentally observed
scaling laws characteristic to BN.

\section{The spring-block model}

\begin{figure}
\centering
\epsfig{figure=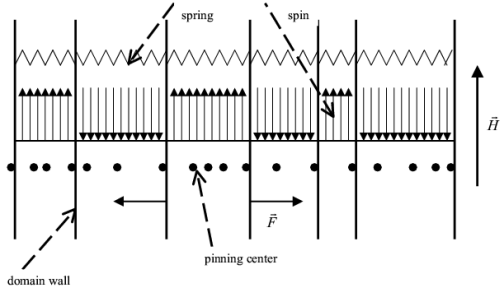, angle=0, width=3.0in}
\caption{Sketch of the mechanical spring-block model}
\label{rajz}
\end{figure}

The model is essentially a one-dimensional spring-block system. It is aimed to reproduce
the accepted microscopic picture of domain wall dynamics for 180 degree Bloch-walls which
separate inversely oriented ($+ | - | + | - | + \dots$) magnetic domains (Fig.\ref{rajz}).

We assume that the domain walls are pinned by defects and impurities, and cannot move unless
the resultant force acting on them is bigger than the strength of the $F_p$ pinning force. When
the resulting force is greater than the pinning force, the wall simply jumps in the resulting
force direction on the next pinning center. Apart of this pinning force there are two other
types of forces acting on each domain wall. To understand these forces let us consider the
$i$-th wall (which separates the $(i-1)$-th and $i$-th domain) free to move and all other walls
fixed. One of the forces acting on the domain wall, $F_H$ , results from the magnetic energy of
the domains $i$ and $(i-1)$ in an external magnetic field. Let us consider the external magnetic
field as sketched in Fig.\ref{rajz}. The interaction energy between one magnetic domain and the
external magnetic field:
\begin{equation}
\label{w}
W=-c_H\cdot H\cdot M
\end{equation}
where $c_H$ is a constant, $H$ is the strength of the external magnetic field and $M$ is the
magnetization of the domain (the positive direction both for $M$ and $H$ is taken upwards).
Taking into account that the $(i-1)$ and $i$ neighboring domains are oppositely oriented, their
total energy of interaction with the external magnetic field is:
\begin{equation}
W(i)=W_{i-1}+W_i=-c_H\cdot H\cdot \Delta M
\label{dw}
\end{equation}
The quantity $\Delta M$ (the sum of magnetizations of the neighboring $(i-1)$ and $i$ domains) is related
to the two domains' length's difference ($\Delta x=x_i-x_{i-1}$) as
\begin{equation}
\label{dm}
\Delta M={(-1)}^{i}\gamma \cdot \Delta x
\end{equation}
where $\gamma$ is constant relating the size of the domain with it's magnetization. From (\ref{dw}) and (\ref{dm})
it results:
\begin{equation}
\label{wi}
W(i)={(-1)}^{i+1}\cdot c_H\gamma \cdot H\cdot \Delta x
\end{equation}
The force $F_H$  acting on domain wall $i$ can be determined considering the $\delta L$ elementary work
performed by this force, when the wall is displaced by a distance $dl$
\begin{eqnarray}
\label{dl}
\nonumber \delta L=F_H\cdot dl=-dW(i)={(-1)}^{i}c_H\gamma\cdot H\cdot d(\Delta x)= \\
={(-1)}^{i}c_H\gamma\cdot H\cdot 2dl
\end{eqnarray}
\begin{equation}
\label{FH}
F_H=\frac{\delta L}{dl}={(-1)}^{i}2c_H\gamma \cdot H={(-1)}^{i}\beta \cdot H
\end{equation}
with $\beta=2c_H\gamma$ another constant. In our model for the sake of simplicity we define the units such
that $\beta=1$. For positive values of the external magnetic field this force encourages the
increase of the domains oriented in the + direction, and for negative values of the external
magnetic field this force tends to increase the size of the domains oriented in the --
direction.

A second type of force, $F_m$, acting on both sides of the domain walls, is due to the magnetic
self-energy of each domain. This force tends to minimize the length of each domain. It can be
immediately shown that $F_m$ is proportional with the length of the considered domain. The $E_i$
magnetic self-energy of a magnetic domain $i$ has the form
\begin{equation}
\label{ei}
E_i=c_m{M_i}^{2}
\end{equation}
where $c_m$ is a constant. As in the previous case
\begin{equation}
\label{dei}
dE_i=c_m\cdot d({M_i}^{2})=-\delta L=-{F_m}dl
\end{equation}
and from here the $F_m$ force:
\begin{eqnarray}
\label{Fm}
\nonumber F_m=-\frac{dE_i}{dl}=-2c_m(M_i)\frac{dM_i}{dl}\approx -2c_m{\gamma}^{2}\cdot x_i\frac{dx_i}{dl}= \\
=-2c_m{\gamma}^{2}x_i=-{f_m}{x_i}
\end{eqnarray}
The constant $f_m$ is an important coupling parameter in this model and acts as the elastic
constant of some mechanical springs.

The system of the $F_p$, $F_m$ and $F_H$ forces can be now easily mapped on a mechanical spring-block
model.

The main constituents in this mechanical model are randomly distributed pinning centers, rigid
walls sitting on pinning centers (describing Bloch-walls) separating + and -- oriented domains
and springs between the walls (describing the $F_m$ forces). The strength of the pinning centers
(pinning forces), $F_p$, are randomly distributed following a normal distribution. Walls can be
only on pinning centers and two walls are not allowed to occupy the same pinning center.
This constraint implies that the number of magnetic domains and domain walls are kept constant
and are thus a-priori fixed. Domains cannot totally disappear and new domains cannot appear
during magnetization phenomena. The elastic springs are ideal with zero equilibrium length
and with the tension linearly proportional with their length. The tension in the elastic
springs will reproduce the $F_m$ forces. Beside the pinning forces and the tensions in the springs
there is an extra force acting on each wall. The strength of this force is proportional
with the exterior magnetic fields intensity, it is the same for all walls but its direction is
inverse for $+|-$ and $-|+$ walls. This force will reproduce the $F_H$ forces.

The dynamics of this model is aimed to reproduce real magnetization phenomena. First $N_p$
pinning centers are randomly distributed on a fixed length ($L$) interval, and their strengths
are assigned. Than a fixed $N_w$ number of walls are randomly spread over the pinning centers
($N_w \ll N_p$) and connected by ideal springs. Neighboring domains are assigned opposite magnetic
orientation. The external $F_H$ force is first chosen zero (corresponding to $H=0$), and we let the
system relax to an equilibrium configuration. To achieve this we calculate the resultant of
the $F_m$ forces on each wall. If the strength of the pinning force acting on one wall is smaller
than the force resulting from the tension of the springs attached to it, the wall will jump in
the direction of the resultant force on the next pinning site. However, if this pinning site
is already occupied, the wall remains in its original position. We assume that the time needed
for the system to achieve equilibrium is zero. It is important to note that one event (jump)
can trigger many other events leading to avalanche-like processes. The above dynamics is
continued until the equilibrium is satisfied for each domain wall. The order in which the
position of the walls is updated is random.

Once the initial equilibrium configuration is reached we begin to simulate the magnetization
phenomenon. The value of the $F_H$ external force is increased step-by-step (corresponding to an
increasing $H$ magnetic field intensity), and for each new $F_H$ value an equilibrium position of
the system is searched. In equilibrium the magnitude of the resulting force on each wall should be
less than the pinning force acting on that wall.
In each equilibrium configuration we calculate the total magnetization of the system as:
\begin{equation}
\label{m}
M=\sum_{i}l_i\cdot s_i
\end{equation}
where $l_i$ is the length of domain $i$, and $s_i$ is it's orientation: $+1$ for positive
orientation, and $-1$ for negative orientation. We increase $F_H$ until no more walls can move
and the magnetization reaches its maximal value. Starting from this we decrease step-by-step
the value of $F_H$, and for each new value the equilibrium configuration is again reached and
the total magnetization computed. The system is driven until oppositely oriented saturation.
From here we increase again the value of $F_H$ and many hysteresis cycles are simulated.
Throughout the whole simulation we assume that equilibrium is instanteneously achieved
for any value of the $H$ external magnetic field.

During the simulation we are monitoring the variation of the magnetization focusing on the shape
of the hysteresis loop, jump size distribution, power spectrum, Barkhausen signal  duration
and signal area distribution functions.

The hysteresis loop is the history-dependent relation between the magnetization $M$ and the
external magnetic field $H$ when the value of $H$ is increased and decreased successively.
The jump size distribution is the distribution function for the obtained values of abrupt
jumps in $M$ throughout many hysteresis loops. The Barkhausen signal is given by $\frac{dM}{dt}$
and it is proportional with an electric voltage that would be induced in a detecting  electromagnetic
coil. Since there is no real time in our simulations, and the evolution of the system is
related solely to the driving rate ($dH$) of the external magnetic field, we consider a Barkhausen
jump as $\frac{dM}{dH}$, that is the change (jump) in the total magnetization when the driving
field changes by one $dH$ step.

We determined the power spectrum of the obtained Barkhausen signal by using a Fast Fourier
Transform (FFT). Since in our simulations there is no real time, the frequency is also not
realistic, and it is defined solely by $dH$ which can be considered in a first approximation
to be constant in time. Thus, we can emphasize in advance that the power spectrum determined
from this simulation should not be considered relevant.

We also study the shape of the histograms for Barkhausen signal duration and signal area.
In terms of our simulation the signal duration measures the number of consecutive $dH$ steps
when Barkhausen jumps occur ($\Delta M/dH$ is nonzero). Signal area is also related to this quantity:
it represents the area under $\Delta M/dH$ versus $H$ for a nonzero $\Delta M/dH$ sequence.

The  parameters of the model are: $N_p$ -- the number of pinning centers; $N_w$ -- the number of Bloch-walls
($N_w \ll N_p$); $f_m$ -- the coupling constant between the neighboring domain walls (corresponds to the elastic
constant in the case of coupled springs); and the $dH$ -- driving rate of the external magnetic field (change in
$H$ for one simulation step). The total length of the magnetic domains is considered to be of unit length ($L=1$) and the distribution
function for the strength of the pinning forces was considered to be normally distributed and this distribution has been fixed.
We use rigid boundary conditions: the first Bloch-wall compulsory occupies the first pinning center, the last
wall occupies the last pinning center, and these bounding walls cannot move. This constraint means that the
geometrical size of our model system doesn't change during the simulation. As we already mentioned the number of
Bloch-walls and magnetic domains are also fixed within this model, domains can shrink or grow, but they cannot
appear or disappear during the simulation.

\section{Effects of the free parameters}

Before presenting the simulation results let us discuss here the expected effects of the free parameters.
We will use adimensional units for the forces, their strengths are determined relatively to the adimensional
strength of the pinning forces. The $f_m$, $\beta$, $H$ and $x$ quantities are defined through equations (\ref{FH})
and (\ref{Fm}) and they are also considered to be adimensional. The pinning forces can have only positive values
and they are normally distributed on the [0,1] interval with mean $\langle F_p \rangle = 0.5$ and standard deviation $\sigma =0.1$.

\emph{a. Influence of $N_p$.} As $N_p$ increases the pinning centers are closer to each other which causes many
small Barkhausen jumps. Concurrently, small number of pinning centers result less but bigger jumps in
magnetization. From the above arguments one can conclude that the value of the parameter $N_p$ influences directly
the shape of the hysteresis loops and the obtained jump size distribution histogram. As $N_p$ increases the
simulation is more and more time consuming, since many small jumps and thus many intermediate equilibrium
positions are possible. This is thus a first factor that limits the value of the used $N_p$ parameter. We have
performed our calculations with $N_p$ between 1000 and 5000.

\emph{b. Influence of $N_w$.} $N_w$ determines the number of magnetic domains in our model system. Large values
of $N_w$ require long computation times since the equilibrium becomes more and more sophisticated. Because $N_w$
should be much smaller than $N_p$, increasing $N_w$ would lead also to large $N_p$ values, which again makes the
simulation technically difficult. In our simulations we considered $N_w = 100$. The parameter $N_w$ determines
how strongly the springs are stressed. Small value of $N_w$ means that the walls are far from each other and the
coupling springs between them are strongly tensioned. In this case the obtained avalanches in wall movements are
usually longer.  The $N_p/N_w$ ratio is one of the most relevant quantities regarding the outcome of the
simulations. We consider the $N_p/N_w$ ratio to be relatively "small" if it is between 10-20 and "large" if it is
greater then 30. For small $N_p/N_w$ ratio springs are not very stressed, thus small number of jumps will occur
during the magnetization process and strong external magnetic field is needed in order to make the walls jump.
After relatively few jumps  saturation is reached and the walls will form "pairs" that can be destroyed only by
inverting the external magnetic field's direction. For large $N_p/N_w$ ratio the existence of the relatively many
pinning centers causes many small jumps. The small number of Bloch-walls causes springs to be stretched and favors
the occurrence of jumps even for weak external magnetic fields. A large number of steps is needed until saturation
is reached and the walls are stopped by their neighbors.  In simulations we varied this parameter between 10 and
50 and studied it's influence on the BN statistics.

\emph{c. Influence of $f_m$.} Since this parameter acts in our spring-block model like the elastic constant of
the springs, its value determines the value of attractive forces between neighboring walls. As $f_m$ increases
the coupling becomes stronger, and weak external magnetic field is enough to make the walls jump. For small $f_m$
values the springs are weakly coupled, so a stronger external magnetic field is needed to make the walls jump.
In our model the $N_p/N_w$ ratio and the $f_m$ parameters are strongly related to each other. In the case of many
pinning centers and relatively small number of walls (equivalent with $N_p/N_w$ large) even for weak coupling
($f_m$ around 10) many jumps occur and equilibrium states are easily reached. When the $N_p/N_w$ ratio is small
(around 10) weak coupling requires strong external magnetic fields in order to make walls jump and only a
relatively small number of jumps are possible. In this parameter region the hysteresis loops have only a very limited
number of jumps and these jumps occur only for high $H$ magnetic field values. If the coupling gets stronger
(still $N_p/N_w$ low) equilibrium states are very difficult to reach, walls jump back and forth and hysteresis
loops are totally damaged. When the $N_p/N_w$ ratio is large (around 30), this means that there are many pinning
centers and relatively small number of walls. In this case weak coupling ($f_m$ around 10) will be enough to make
walls jump even for low $H$ values because the springs are strongly stressed. The expected result is the
existence of many small jumps along the whole hysteresis loop. For stronger couplings equilibrium is again
difficult to reach and the hysteresis curves are expected to be damaged. It is obvious that finding the optimal
parameter region for the simulation is crucial. Wrong parameters that generate damaged hysteresis loops, or
situations  where the equilibrium states are difficult to reach (walls jump back and forth many times consecutively)
will cause significant artifacts in the jump-size, signal-duration or signal-area distribution functions.

\emph{e. Influence of the driving rate in $H$ ($dH$).}  This parameter is also important for all the considered
distribution functions. It is obvious that small $dH$ (around 0.001) steps produce very many "short" and small
jumps, while larger steps ($dH$ between 0.01-0.005) make possible only the "longer" Barkhausen-signals. The
latter means that larger $dH$ steps lead usually to many consecutive jumps (i.e. large avalanches). Very small
$dH$ value will divide the bigger and longer jumps into many tinny jumps. In the experiments greater $dH$ steps
correspond to the case where the ferromagnetic sample is submitted to a relatively fast changing driving field,
thus it reaches the saturation magnetization after few number of large jumps. In the experiments it is reported
however \cite{spas,obrien,cote,bahi} that they used very low frequencies ($< 1$ Hz) for the driving field.
This is the reason why we have chosen to make the simulations with relatively small $dH$ steps (0.001) that
corresponds to quasi-stationary driving and allows the system to relax to a closer equilibrium
configuration during each step of the
magnetization -- demagnetization process.

Pondering the effects of all the above described parameters, we have found that the best parameter region that
produces simulation results in agreement with the experimental ones is the following: $N_p = 1000-5000$,
$N_w = 100$ ($N_p/N_w = 10-50$), $f_m=10$, $dH=0.001$. The simulation results presented in this paper are all
obtained for this parameter region. The relevant distribution functions were obtained by averaging on $10$ independent configurations (different initial states for the walls and pinning centers) and considering $10$ hysteresis loops for every configuration.

\section{Simulation Results}

\begin{figure}
\epsfig{figure=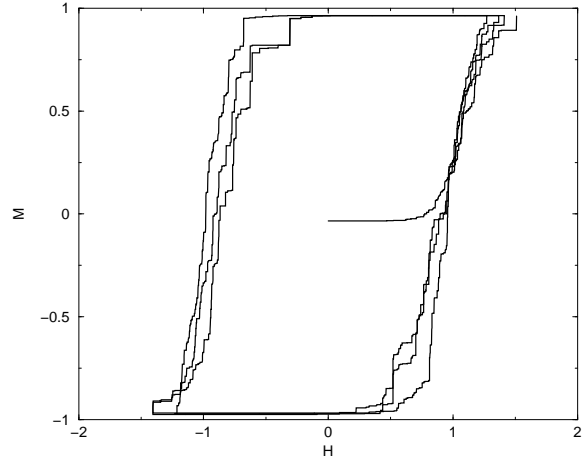, angle=270, width=3.0in}
\caption{Hysteresis loops obtained by simulation with parameter values: $N_p=3000$, $N_w=100$,
$f_m=10$ and $dH=0.001$.}
\label{hyst}
\end{figure}

Characteristic hysteresis loops are plotted on Fig. \ref{hyst}. The shape of the obtained hysteresis curves satisfies
our expectations and fulfills all the requirements for real magnetization phenomena. On these curves one can
detect many discrete jumps with different sizes, thus the model exhibits BN. In addition, when the sample is
driven consecutively through many hysteresis cycles the magnetization curves do not follow exactly the same
path, although the parameters of the simulation were unchanged. The qualitative shape of the hysteresis curve is
quite stable for a wide range of the free parameters.

\begin{figure}
\epsfig{figure=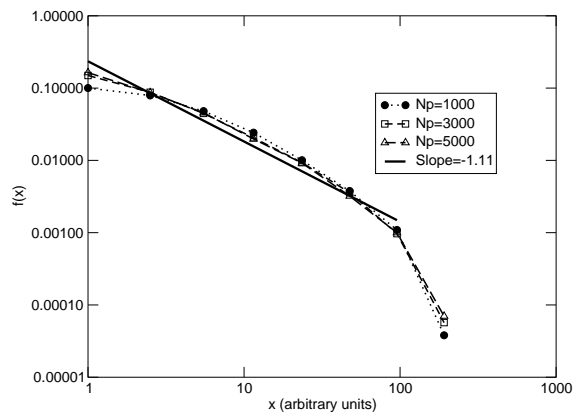, angle=0, width=3.5in}
\caption{Jump size distribution function for $N_p=$1000, 3000 and
5000, $N_w=100$, $f_m=10$, $dH=0.001$. The solid line indicates a
power-function with slope = -1.11.}
\label{jsd}
\end{figure}

The jump size distribution histogram in our simulations corresponds to the avalanche size distribution histogram
in experiments and it is the most relevant distribution for the characterization of the Barkhausen noise. For
$N_w=100$, $N_p$=1000, 3000 and 5000, $f_m=10$ and $dH=0.001$, the jump-size distribution function

\begin{equation}
\label{fx}
f(x)=\frac{N(x,x+dx)}{N_t\cdot dx}
\end{equation}
is plotted on Fig. \ref{jsd}. ($N(x,x+dx)$ is the number of magnetization jumps with sizes
between $x$ and $x+dx$, and $N_t$ is the total number of jumps during the measurement).

Based on this graph we
can conclude: (i) for small values of $N_p$ the histograms does not show a clear power-law distribution for the
jump sizes; (ii) the curves for increasing $N_p$ numbers suggest however that for larger system sizes the jump size
distribution approaches a power-law; (iii) taking into account this "straightening" tendency for increasing $N_p$
values we have fitted a power-law function with slope $=-1.11$ as a guideline. The region where this scaling is
valid extends however to only one decade in our simulations and represents the trend that is to be followed if
$N_p$ is increased to infinity; (iv) the cutoffs at the ends of the histograms seem to be finite-size effects.

The fact that our simulation results indicate a power-law distribution in the thermodynamic limit means that
jump (avalanche) sizes show scale-invariance, a feature that is expected to be common for all types of crackling
noises.

\begin{figure}
\epsfig{figure=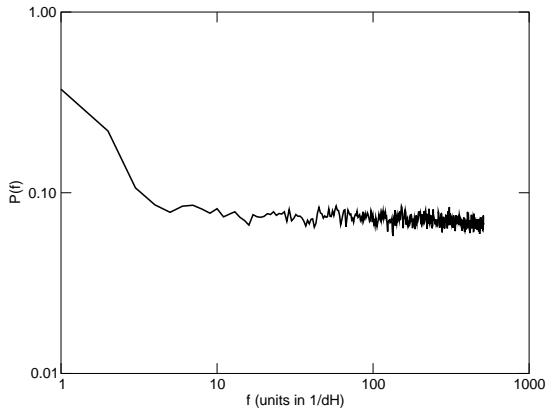, angle=0, width=3.5in}
\caption{Power spectrum of the simulated BN ($N_p=3000$; $N_w=100$; $f_m=10$).}
\label{fft}
\end{figure}

The power spectrum obtained from simulation suggests a white noise over two decades (Fig. \ref{fft}.). We emphasize
it again that this is not a real power spectrum since we do not have real time in simulations, and thus we cannot
define frequency. Time evolution is substituted with the driving rate $dH$ of the external field and it is
considered that equilibrium is reached for each simulation step. We have plotted thus the "power spectrum" in
terms of the $1/dH$ -- type "frequency", assuming the unit time as the time needed to change the external magnetic
field by $dH$.

\begin{figure}
\epsfig{figure=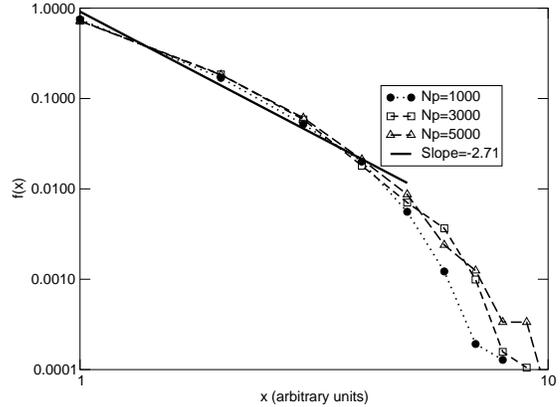, angle=0, width=3.5in}
\caption{Distribution function for signal duration ($N_w=100$;
$f_m=10$; $dH=0.001$). The solid line indicates a power-function with slope$=-2.71$}
\label{sdur}
\end{figure}

On Fig. \ref{sdur}. we plotted the signal duration distribution functions for different system sizes ($1000-5000$). As it
results from the graph in our simulations the longest jumps lasted around 10 $dH$ steps, we had thus only one
decade of data. However, the tendency is obvious again: the larger the system is, the more the distribution function
tends to reach a power-law. We have plotted as a guideline the power-function with slope$=-2.71$ that fits the
central part of the $N_p=5000$ histogram. The reason that only relatively "short" jumps occurred in our
simulation is that the driving was chosen to be quasi-static. The value of $dH$ was chosen small (0.001) and thus
the jumps are small and short. Most of the time one $dH$ step is not enough to provide the needed
amount of energy for a further jump if in the previous step a jump occurred, thus the big majority of Barkhausen
jumps lasted only for one $dH$ unit.

\begin{figure}
\epsfig{figure=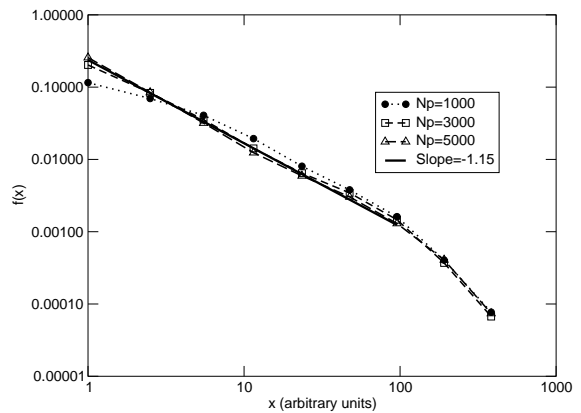, angle=0, width=3.5in}
\caption{Signal area distribution function. For the specified $N_p$ values we used $N_w=100$; $f_m=10$; $dH=0.001$. The solid line is a power-function with slope$=-1.15$.}
\label{sarea}
\end{figure}

On Fig. \ref{sarea}. we plotted the signal area distribution function for increasing system sizes ($N_p=1000-5000$). The
curves suggest a clear tendency: the larger the simulated system is, the better a power-law approximation is.
From our simulation data we would predict for large systems a signal area distribution function which is a
power-law with exponent around $-1.15$. The cutoffs in our simulations are due to the finite size effects.

\section{Discussion}

Comparison of the obtained simulation results with the experimental ones yields the following conclusions:

1. The simulated hysteresis loops are visually in good agreement with the ones obtained in experiments. The best
curves are obtained with the parameters: $N_p=5000$, $N_w=100$, ($N_p/N_w = 50$), $f_m=10$ and $dH=0.001$. On the
simulated hysteresis loops one can observe many Barkhausen jumps with various sizes, just as it is expected from
the experimental results.

2. The shape of the simulated jump size distribution function predicts in the thermodynamic limit
($N_w \rightarrow \infty$, $N_p\rightarrow \infty$, $N_p/N_w \rightarrow \infty$) a tendency towards a power-law. Experimental avalanche size
distribution functions exhibit a power-law behavior with an exponent around $-1$ to $-1.3$ \cite{umm,bahi}. Our results
on relatively small systems suggests a power-law behavior with an exponent $-1.11$. This result is in a reasonable
agreement with experiments and proves that our model is able to account for the most important statistics of the
BN,  although the simulated system's sizes were small and finite size effects were strong.

3. As already emphasized, the power spectrum obtained within our model is not relevant, since we do not have real
time in our simulations. We considered that the equilibration of the system takes place instantaneously and the
detection of the Barkhausen signal is without any inertia.  However, our result which suggests a white noise
(exponent of the power spectrum $\approx 0$) is in agreement with O. Narayan's  \cite{nara} prediction for
one-dimensional systems. Comparing our results with the experiments we found that experiments performed with a
SQUID device \cite{obrien} yield also a power spectrum closer to white noise than $1/f$ noise, which is thus in
agreement with the present simulation results. The very different experimental results (power spectra showing
white noise, $1/f$ noise and even Brownian noise) allow us to conclude that the shape of the power spectrum
depends on the experimental setup and it is not relevant to the Barkhausen noise. Similar conclusions were
suggested in \cite{zent}. It seems that the shape of the experimentally measured power spectra depends on the
inertia of the pickup coil. This may be the reason why experiments performed with SQUID device \cite{obrien}
indicated that BN is closer to a white noise than $1/f$ noise. SQUID devices can follow quickly, without any
inertia the changes in the magnetization and this corresponds to the dynamics involved in our simulations.

4. The experimental results for the signal duration and the signal area \cite{plew,cote} distribution indicated
a power-law behavior for both quantities. For the signal duration distribution the measured exponents were $-2.2$
\cite{plew} and $-1.64$ to $-1.82$ \cite{cote}. Although there is a strong difference between these results,
the experiments agree in the validity of the power-law distribution. Our model also suggests a power-law
distribution in the limit of infinite systems and the predicted exponent is $\approx -2.71$.  For signal area
distribution experiments obtained power-laws with  exponents from $-1.7$ to $-1.8$ \cite{plew} and $-1.74$ to $-1.88$ \cite{cote},
values that are in good agreement with each other. Our simulations predict however a value around $-1.15$. The
numerical agreement is thus not too good, but the power-law tendency suggested in our simulations can explain at
least qualitatively the statistics.

\section{Conclusions}

As a conclusion, we can affirm that the model presented in this paper is able to describe and explain qualitatively
the characteristic features of the BN. The model is realistic and reproduces in a pedagogically simple manner the
microscopic dynamics of the magnetic domain walls. For a quite broad parameter region our simulation results
proved to be at least in qualitative agreement with the known experimental results on the statistics of the BN.
Despite of the encouraging results the model is not perfect. One very important feature is the absence of the
temperature as parameter. Also, there is no real time in simulations, and only the value of the $dH$ magnetization
step determines the rate at which time evolves in our simulations. The model doesn't account for the pinning
mechanism and the strength of the pinning forces. It is an oversimplified one-dimensional approximation for the
complex three-dimensional domain topology. Seemingly the most serious problem of the present model is that the
number of domain walls is a priori fixed and domains cannot appear or disappear during the dynamics. Despite all
these deficiencies this model offers a simple and visual picture of magnetization phenomena reproducing
qualitatively well the statistics of BN.

\section{Acknowledgments}

The work of K. Kov\'acs was supported by the KPI Sapientia foundation.
Z. N\'eda acknowledges a NATO research fellowship, the hospitality and stimulating atmosphere of the Centro de Fisica do Porto.

\end{document}